\documentclass[twoside]{article}
\usepackage[papersize={5.5in,8.27654in},total={4.25in,7.25in},%
  includehead,nofoot,centering]{geometry}
\usepackage{amsmath,txfonts}
\usepackage[T1]{fontenc}
\usepackage[colorlinks=true,allcolors=blue,bookmarks=false,
  pdfauthor={Alphonse Duleau (ed. Gavin R. Putland)},
  pdftitle={Obituary for Augustin Fresnel},
  pdfkeywords={optics, history of physics, history of optics, diffraction, polarization, lighthouses},
  pdfsubject={Optics, Physical Optics, History of Optics, Fresnel}]{hyperref}
\usepackage[final,kerning,spacing]{microtype}
  \SetExtraKerning{encoding={*}}{1={-35,-45}, / ={50,50},
    \textemdash={50,50}, \textendash={30,30}, \textquotedblleft={-30,}}
  \SetExtraSpacing{encoding={*}}{A={-80,,}, f={80,,}, r={-80,,}, v={-80,,},
    w ={-70,,}, y={-60,,}, ' = {-200,,}, \textquotedblright={-80,,}}

\title{{\bf Obituary for Augustin Fresnel}\vspace{-.6ex}}
\author{\large by Alphonse Duleau\\[2ex]\normalsize Annotated English translation of{\tiny\,} Duleau's ``Notice sur A.\,Fresnel'' in\\[-.3ex]\normalsize\textit{Revue encyclop\'{e}dique}, vol.\,39, pp.\,558--67 (September 1828),\\[-.3ex]\normalsize and of the shorter obituary for Fresnel in \textit{id.}, vol.\,37, pp.\,316--7.\\[1ex]\large Edited by Gavin R.{\small~}Putland~\!\thanks{\,\small Royal Melbourne Institute of Technology, Australia.~ Gmail name: grputland.\\Copyright license: \href{https://creativecommons.org/licenses/by/4.0/legalcode}{Creative Commons Attribution 4.0 International}.}}
\date{\vspace{-1ex}\normalsize 3 March 2026} 

\makeatletter
\def\ps@plain{
  \let\@oddfoot=\@empty
  \def\@oddhead{\hfill\normalsize\sf\thepage}
  }
\def\ps@headings{
  \def\sectionmark##1{\markright{\S\thesection.~ ##1}}
  \let\@evenfoot=\@empty  \let\@oddfoot=\@empty
  \def\@evenhead{\footnotesize\sf\underline{\makebox[\textwidth]{\normalsize\sf\thepage\hfill\footnotesize\sf\phantom{g}Duleau (ed.~Putland),\itshape\, Obituary for Augustin Fresnel}}}
  \def\@oddhead{\footnotesize\sf\underline{\makebox[\textwidth]{\rightmark\phantom{g}\hfill\normalsize\sf\thepage}}}
  }
\makeatother

\begin{document}

\sloppy

\maketitle

\pagestyle{headings}

\vspace{-1em}

\section*{Editor's preface}\label{s-preface}

Alphonse Jean Claude \textsc{Bourguignon}, called \textsc{Duleau}, was
born in Paris on 30~April 1789~\cite{anon-1833}, and entered the
\textit{\'{E}cole des ponts et chauss\'{e}es} (School of Bridges and
Roads) in~1809. As a student engineer in~1810 he was assigned to
Pontivy, where he became acquainted with Augustin
Fresnel~\cite{pernot-23}. After graduating in 1812, Duleau was sent to
Bordeaux and the Garonne district to work on bridges, and finally back
to Paris~\cite{PIREN-Seine-26}, where he worked on the completion of
the Canal de l'Ourcq and the construction of the Saint-Martin Canal,
including its swing bridges~\cite{anon-1833,pernot-23}. He~rose to
become a professor of construction, a~member of the School's council,
and an editor of the \textit{Annales des ponts et
  chauss\'{e}es}~\cite{PIREN-Seine-26}.

For a proposed forged-iron bridge over the Dordogne at Cubzac, near
Bordeaux, in~1812,\footnote{\,\small Bell~\cite{bell-89} gives the
year as 1811; Duleau himself~\cite{duleau-1820} gives 1812.} Duleau
did his experiments on the ``tension, compression, flexure, torsion,
and elastic stability of full-size bars and structures having
circular, triangular, elliptical, and rectangular cross
sections''~\cite[p.~\!157]{bell-89}, which \emph{preceded} the
theoretical work of Poisson and Cauchy on linear elasticity.

In the third part (``Bibliographic Bulletin'') of the April 1828 issue
of \textit{Revue encyclop\'{e}dique}
(archive.org/details/revueencyclopd38pari), as item~84
(\href{https://archive.org/details/revueencyclopd38pari/page/191}{p.~\!191}),
we find the following:
\begin{quotation}\small 84.\ \textemdash\ * \emph{Notice on
  A.\,Fresnel}, Engineer in Chief of the Royal Corps of Bridges and
  Roads, Member of the Academy of Sciences, etc.; by
  Mr.\,\textsc{Duleau}, Engineer of Bridges and Roads. Paris, 1827;
  Huzard-Courcier.\, Octavo, from one printing sheet.

  We shall insert almost the entirety of this interesting Notice,
  produced only in a small number of copies, in the first part of our
  collection; thus we shall discharge our debt to one of the scholars
  whose loss has most afflicted the friends of the sciences, from
  which \textsc{Fresnel}, still quite young, has been taken in the
  times of the greatest fertility of genius, when the greatest hopes
  were founded on him, and he was ready to realize
  them.~~~\hfill\emph{F.}
\end{quotation}
The description implies, and numerous online sources confirm, that the
original \textit{Notice} was a 16-page brochure. But, as the full text
of the brochure does not seem to be available, I~can only hope that
``almost'' the entirety means the entirety of the substantive
content. Entire or otherwise, the promised reprint~\cite{duleau-1828}
appeared in the September 1828 issue of the \textit{Revue}. It~was
reprinted again, in the last quarter of the same year, in the
inaugural volume of \textit{Journal du g\'{e}nie civil, des sciences
  et des arts} (Journal of Civil Engineering, Sciences, and Arts),
with an added prefatory paragraph:
\begin{quotation}\small \emph{N.B.}\, We borrow this notice from the
  \textit{Revue encyclop\'{e}dique}; it~is so outstanding that our
  collaborator who was charged with similar work thought, as we did,
  that one could not do better than to reprint~it. To~do so is,
  moreover, a tribute paid to the remarkable talent of the learned
  engineer who drafted it, and a mark of esteem for the zeal and skill
  of the editor of the \textit{Revue}, who for ten years has not
  ceased to render such important services to the sciences, the arts,
  and all human knowledge.
\end{quotation}
In~the present translation, unsigned footnotes are in the original (or
at least the reprints thereof), but the numbering of footnotes has
changed (the old numbering was within-the-page). Added footnotes are
signed ``\emph{Ed.}''\, Other editorial additions, including citations
and original page numbers, are enclosed in square brackets; e.g.,
``[559:]''~marks the beginning of p.~\!559 in the first reprint. A
translation of a shorter obituary~\cite{a-1828} cited in the reprints
of Duleau's obituary (footnote~\ref{f-shorter} below) is included as
an \hyperref[s-appx]{appendix}.

Volume~52 of the \textit{Annales de chimie et de~physique} (first
quarter, 1833) included obituaries for 28~persons, nine of whom were
reported to have died of cholera. One of the nine was Duleau, who was
taken on 26~April 1832, aged not~quite~43, having fallen ill while
leading construction of some of the sewers of
Paris~\cite{anon-1833}.\, A more detailed obituary on him, in the
\textit{Compte-rendu trimestriel} of~the \textit{Association
  polytechnique}, was written by L\'{e}onor
Fresnel~\cite{fresnel-l-1833}.

\bigskip

\begin{center}

\section*{\LARGE\sc{\bfseries Notice~ on~ A.~Fresnel}{\tiny\,}\large,}

\large\emph{Engineer in Chief of the Royal Corps of Bridges and Roads,
Member of the Institute, of the Philomatic Society, of the Royal
Society of London, of the Society of Physics and Natural History of
Geneva, of the Royal Academy of Caen, Knight of the Legion of Honor;
died at Ville-d'Avray, 14~July~1827.}\footnote{\,\small See
\textit{Rev.\,Enc.}, vol.\,37, p.~\!316
\cite{a-1828}.\label{f-shorter}}

\end{center}

{\small

\textsc{A.\,Fresnel} has been taken away in the prime of life, in the
midst of a career that was already distinguished by important
discoveries and by brilliant and useful applications, and which
promised a far more generous harvest of glory. His premature death is
an irreparable loss, which has afflicted the friends of science no
less than his own family.

His worthy friend, Mr.\,\textsc{Arago}, having given the most touching
farewell at his grave, will soon set forth Fresnel's claims to public
recognition and admiration~\cite{arago-1830}. It belongs to him, who
was privy to his genius and his virtues, to show them in all their
brilliance. May I be permitted to precede Mr.\,Arago in this further
tribute to the memory of our mutual friend. Mr.\,Arago's words will be
addressed more especially to the learned world. For my part, having
had the good fortune to be admitted into Fresnel's inner circle, and
to live, so to speak, as one of his family, I~have no other aim than
[559:]~to converse for a moment with his friends, about the one whom
they will never forget.\medskip

}

\emph{Augustin-Jean} \textsc{Fresnel} was born at Broglie, in the
d\'{e}partement of Eure, on 10 May 1788. At the age of sixteen he
entered the \textit{\'{E}cole polytechnique}, where his elder
brother\footnote{\,\small The eldest [Louis,
  \href{https://gw.geneanet.org/pierfit?lang=en&n=fresnel&p=louis}{1786--1809}],
having volunteered to join the artillery, died in Spain in
1807~[\href{https://gw.geneanet.org/pierfit?lang=en&n=fresnel&p=louis}{sic}].
The third-born [L\'{e}onor, 1790--1869] is an engineer of Bridges and
Roads. A fourth [Fulgence, 1795--1855] is successfully pursuing a
literary career.} had preceded him and where his younger brother
followed him. There all three shone in the first rank.

A feeble constitution and delicate health perhaps favored Augustin's
inclination for observation and reflection. This valuable disposition
in him was united with a fine and delicate mind, a sincere modesty, a
benevolent cheerfulness, and an unalterable gentleness\textemdash
qualities which took their source in sound judgment and a pure heart,
and which foretold the solid virtues which important occasions would
put to the test.

On leaving the \textit{\'{E}cole polytechnique}, Fresnel embraced the
career of Bridges and Roads. First called to the d\'{e}partement of
Vend\'{e}e, which the government was covering with roads and
constructions as if improvised, then to the d\'{e}partements of
Dr\^{o}me and Ille-et-Vilaine, he found himself for several years
almost entirely absorbed by public service. Only his colleagues could
appreciate the conscientious care with which he probed the minutest
details of his work, placing these duties before all, and giving way
only in secret, and for relaxation, to research and reflection on
mathematics and physics.

The political events of 1814 \& 1815 gave him, amid the general
upheaval, some leisure to continue his scientific work. But before
speaking of him as a scientist, let us consider him as a citizen, in
the midst of [560:]~the unrest that broke out in the south of France
at the beginning of~1815.\footnote{\,\small That is, when Napoleon
returned from Elba.~\textemdash\emph{Ed.}} A friend of wise liberty,
he regarded representative government, established by the Charter, as
most suitable for an enlightened people. An unforeseen storm having
fallen on this barely-established government, Fresnel believed that
every Frenchman must devote arm and life to the defense of the
constitutional monarchy. He was then at Nyons; he joined the royalist
army of the south, despite the dangers to which he was exposed by the
intoxication of enthusiasts fired up by the still recent memories of
the Empire. This step was without effect. Never afterwards did he seek
to capitalize on this act of devotion, which the ruined state of his
health made doubly meritorious: he believed he had only done his duty.

It was at Nyons that Fresnel made his first observations on the novel
phenomena presented to him by \emph{diffraction of light}\textemdash
phenomena inexplicable under the \emph{emission} theory that the
greater number of the learned, following Newton, had adopted. These
facts, on the contrary, agreed with the \emph{theory of Huygens and
Euler, which attributes the phenomena of light to vibrations of a
fluid spread through all space}.

The memoir in which Fresnel's first researches are set
out~\cite{fresnel-1815} was submitted in~1815 to the Academy of
Sciences. In~the isolation in which he had found himself until then,
he was unaware that a learned Englishman, the celebrated
Dr.\,Thomas{\tiny\!} Young, occupied with the same phenomena, had
already arrived at some results that had furnished potent arguments in
favor of the \emph{theory of undulations} [wave theory]. Some time
afterwards the Academy proposed, as the subject of a prize, the
general examination of the \emph{phenomena of diffraction}. Fresnel
pressed on much further with his discoveries~\cite{fresnel-1818mc}
and, in~1819, won the proposed prize.

On his return to Paris in~1815, Fresnel had the good fortune to come
into contact with Mr.\,Arago.\footnote{\,\small The meeting in 1815
preceded Fresnel's first memoir on
diffraction~\cite[pp.\,594--5]{boutry-48}. But Fresnel was again in
Paris from March to October of
1816~\cite[pp.\,597,\,599]{boutry-48}.~\textemdash\emph{Ed.}} The
ardor of that celebrated academician for all that concerns the
progress of science initiated their connection; a~conformity of
studies, [561:]~tastes, and principles cemented it. Fresnel owed to
this friendship a more rapid development of his discoveries. Modest,
shy, perhaps even a little mistrustful [\textit{d\'{e}fiant}], he
needed, as a confidant in his research, a~scientist with whom he could
express himself without reserve; he found in Mr.\,Arago a friend who
knew how to procure for him every means of research, to study nature
with him, to appreciate and bring to light his discoveries: in~short,
to reveal his genius to others and to himself.

Mr.\,Becquey, Director-General of Bridges and Roads, by placing
Fresnel in Paris, gave him the means to continue his scientific
labors. This administrator, who has always accorded enlightened
protection to the sciences and their applications, and whose paternal
solicitude for the corps that he directs is known to all, had
recognized the genius and noble character of the young physicist, and
granted him the esteem and affection of which he was worthy.

My aim here is not to set out what Fresnel did to deduce, from the
single fruitful principle whose validity he had proved, the various
phenomena of light, each of which, under the emission theory, required
a particular hypothesis which was, so to speak, only a translation of
the observed fact. I~must confine myself to saying that with the aid
of this principle, he fully explained \emph{diffraction, inflexion,
reflection, polarization, refraction, double refraction,
etc.}\,\cite{putland-2021-}; in the new theory all these phenomena,
until then largely independent of one another, form a system in which
everything is linked by the simplest mechanical considerations. In
several of his memoirs, all the resources of geometry and analysis are
applied to the most delicate and most ingenious experiments. The
researches to which he was still devoted in the last months of his
life had as their object \emph{the difference in dispersive power of
various media, compared with their refractive power}.

Works so remarkable earned him, in~1819, his admission to the
Philomatic Society; in~1823, his reception into the Royal Academy of
Sciences by a unanimous vote, [562:]~an honor that very few scientists
had obtained before him; in~1824, the decoration of the Legion of
Honor; in~1825, admission to the Royal Society of London; and finally,
in~1827, for having applied the \emph{theory of undulations} to the
phenomena of
\emph{polarization}~\cite{fresnel-1821c-putland,fresnel-1821d-putland,fresnel-1822z-putland},
the prize of the Royal Society of London endowed by Count Rumford for
the finest discovery on the theory of heat and light.

At the same time as he was devoted to the most fundamental researches
on the theory of light, he was led to make a brilliant application of
the principles of optics to the lighting of lighthouses. The immense
advantage that navigation draws therefrom ranks this invention among
the most useful services rendered to humanity. Called in~1819 to be a
member of the Lighthouse Commission, and appointed by Mr.\,Becquey,
Director-General of Bridges and Roads, to do experiments on lighting
together with Messrs.{\tiny\!}~Arago and Mathieu, Fresnel at first
thought of substituting large glass lenses for parabolic
reflectors. This idea, which easily presents itself, had already
become a reality in England; but the lighthouse in which it had been
applied produced little effect, probably because of the great
thickness of the glass. [Count] Buffon had imagined dividing the
convex surface of a lens into several annular zones, and displacing
each of them parallel to the axis so that the rings would have only a
small thickness; but he supposed this stepped surface to be worked
from a single piece of glass, and the execution presented almost
insurmountable difficulties. Fresnel, who had the same idea,
found\textemdash which no one before him had found\textemdash the
means of making it feasible, in having the rings of the same stepped
lens executed separately. This division also furnished the means of
almost entirely correcting the spherical aberration, by giving the
surface of each ring a suitable curvature.

Lenses, thus perfected, transmit nine-tenths of the incident rays,
whereas reflectors return only half. Lenses spread the light much less
than [563:] reflecting surfaces, which, over a large part of their
extent, are very close to the luminous body in relation to its
size.\footnote{\,\small In principle, if not in practice at the time,
one could make the reflector bigger and proportionally further from
the light source, in order to reduce the angular size of the image of
the source at infinity. But this would not overcome the other
objections.~\textemdash\emph{Ed.}} By means of a system of lenses
surrounding the luminous body, Fresnel managed to utilize most of the
rays of this central light\textemdash a result he could not have
obtained using reflectors, which allow the rays emitted directly
through their opening to escape without utility for the distant
observer.\footnote{\,\small That is, without being focused into a
narrow beam visible at large distances. Mirrors, unlike lens panels,
cannot completely surround the source; there must be gaps through
which the reflected beams can escape, and through which unfocused
light radiated directly from the source also
escapes.~\textemdash\emph{Ed.}} Another advantage, which alone would
suffice to ensure a marked preference for the use of lenticular
glasses [i.e.,~lenses], is the inalterability of their surface;
metallic reflectors, on the contrary, need frequent cleaning, which
impairs their form and their lustre.

The central flame for the polygonal layout of lenses that compose a
lighthouse needed to be at once vivid and compact.\,
Messrs.{\tiny\!}~\emph{Arago} and \emph{Fresnel} solved this problem
by following the idea of Count \emph{Rumford} on multiple
nozzles~\cite[p.~\!11]{fresnel-1822i-tag}, and by completely
eliminating the grave inconveniences of the high temperature by
applying the ingenious invention of \emph{Carcel}, which consists in
supplying the wicks with an overflow of
oil~\cite[pp.\,11,~\!18]{fresnel-1822i-tag}.

Fresnel at first had a revolving first-order light built, composed of
eight large lenses positioned in an octagonal prism: each of them is
equivalent, in \emph{total effect}, to three of the largest reflectors
hitherto employed.\footnote{\,\small A lamp with four concentric
wicks, lit at the focus of one of these large lenses, has been seen
with the naked eye one hour after sunset, at a distance of 50 English
miles or 15 nautical leagues, and appeared as bright as an English
light at only a third of that distance. (\textit{Geodetic operations
  done in autumn} 1821, by Messrs.{\tiny\!}~\textsc{Arago} and
\textsc{Mathieu}.)}

A lighthouse of this kind has been in operation since 1825~[sic] at
the mouth of the Gironde, on the tower of Cordouan.\footnote{\,\small
The exact date, according to innumerable sources, is 25~July
182\emph{3}{\tiny\,}\textemdash the bicentenary of which was much
publicized on French websites. However, the ``system'' mentioned later
in the paragraph was indeed unveiled in 1825.~\textemdash\emph{Ed.}}
Two small fixed-light lighthouses composed of \emph{cylindrical
lenses} have been placed, one at Dunkirk, the other at the Pointe de
Grave. These three lighthouses form part of a complete system adopted
by the Director-[564:]General of Bridges and Roads, on the advice of
the Lighthouse Commission,\footnote{\,\small See the \textit{Report by
  Rear-Admiral de {\sc Rossel} on the new lighthouse lighting system
  of France}. In~England, Holland, Denmark, Russia, Tuscany, etc., the
system of lenticular lighthouses was soon appreciated.\,
Mr.\,\textsc{Stevenson}, the famous Scottish engineer, having
inspected the beautiful lighthouse of Cordouan in great detail,
requested and obtained from Mr.\,\textsc{Becquey}, Director-General of
Bridges and Roads, that two large lenses and a first-order lamp be
placed at the disposal of the Scottish Lighthouse Commission.} for
lighting the coasts of France.

Mr.{\tiny\!}~de \textsc{Chabrol}, Prefect of the Seine, always eager
to foster the development of useful discoveries, thought that the
lenticular system with some modifications could be applied to the
lighting of the quays of the Saint-Martin Canal, and engaged Fresnel
to conduct research on this subject. The scientist, to satisfy the
complex conditions of the proposed program, devised a catadioptric
apparatus which he was not able to see completely finished. The trials
done on it very recently have given satisfactory results, and make us
hope that with slight modifications, this system will be able to find
the most useful application in the lighting of the streets and public
squares. The same system, simplified in some respects by its author,
will be applied very advantageously to lighting the entrances of
seaports.

At the exhibitions of 1823 and 1827, one could see the lenticular
systems of \textsc{Fresnel} executed by
Messrs.{\tiny\!}~\textsc{Soleil}, optician, and
\textsc{Tabouret},\footnote{\,\small \textsc{Fresnel} discovered in
this young man, still a stranger to the first elements of the
sciences, a~rare aptitude for the mechanical arts. At the same time
that the physicist was directing Mr.\,\emph{Soleil}, he was training
Mr.\,\emph{Tabouret}, who soon afterwards produced lenticular devices
whose precision would do honor to the most able opticians. The
examining jury of the 1827 exhibition has awarded Mr.\,Tabouret a
bronze medal for the execution of various devices.} a driver in the
Bridges and Roads service. The examining jury of 1823 awarded a silver
medal to Mr.\,Soleil [565:]~for the execution of the lenticular
lighthouses, and placed their inventor outside the competition by
awarding him a gold medal and recommending him for the Order
[literally ``cross''] of Saint-Michel.

The Society for Encouragement [of National Industry], in~1824, awarded
\textsc{Fresnel} the gold medal which it annually assigns to the
author of the finest discovery in the arts.

Reflecting on the manner in which Fresnel became engaged in the work
on lighting of lighthouses, one will be struck by the ease and
fertility of invention that he showed in this situation, where,
receiving the order to occupy himself with a question in many ways new
to him, he responds at once with important discoveries; it has been
truly said that \emph{this is the first time a discovery has been made
by order}. One will also appreciate the recognition owed to the
administrator who foresaw what one could expect from a man like
Fresnel.

The invention and construction of annular lenses will not only have
been useful for lighthouse illumination, but will also serve the
advancement of physics; it provides a potent instrument with which one
can focus the solar rays upon bodies enclosed in a glass vessel, in
order to melt or vaporize them in air or in any other gas. To this new
series of experiments we shall perhaps owe discoveries of high
importance. It was with the aid of this apparatus that Fresnel
discovered the remarkable actions which two unequally heated bodies
exert upon each other at a distance~\cite{fresnel-1825h}.

While thus engaged in learned researches and ingenious applications,
he still found time to attend to other labors. For several years he
was attached to the cadastre of the paving of Paris; he fulfilled this
arduous duty with the most scrupulous exactness.  In~1821 he was
appointed examiner in physics and descriptive geometry at the
\textit{\'{E}cole polytechnique}, a post which [\'{E}tienne-Louis]
\textsc{Malus}, celebrated like him for discoveries on light, had
occupied for several years before him. A premature death, and even the
circumstances of that death, made their destinies [566:]~painfully
similar. Malus and Fresnel, sixteen years apart, both received from
the Royal Society of London, very shortly before their last moments,
the gold medal endowed by Count Rumford.  To~say that Fresnel
distinguished himself by the scrupulous care, the impartial justice,
the prompt and sure sagacity of his predecessor is to pay him, as an
examiner, the highest praise. Feeling all the gravity of the mission
with which he was charged, he applied to it all of his faculties, and
more than his faculties.

In~1824, following one examination at the \textit{\'{E}cole
  polytechnique}, he suffered alarming symptoms indicating an almost
complete exhaustion. From then on he was always fatigued; and after
several fluctuations that left him successively weaker, he has
succumbed in the arms of a family whose happiness and glory he
made.\smallskip

It is for me, the friend of his brother L\'{e}onor at college and at
the \textit{\'{E}cole polytechnique}, for me who owes to that
friendship the most precious happiness of my life, the affection of
Fresnel and the union of our families, for me whom the loss of the one
we regarded almost as a father plunges into the same affliction, who
was present at his last moments, and who gathered his last words,
to~say what were his severe and invariable principles: his reverence
for virtue, which he placed far above science and genius; his strength
of soul, I~will say not only against death, but against the
interruption of discoveries that he had prepared and outlined, and of
which he hoped to derive useful applications. Often, in the last days
of his life, he repeated to me with sorrowful resignation: ``\emph{How
many things I~would still have to do!\tiny\,}'' Some months before his
death we made a long journey together, during which was granted to me
the keenest pleasure, that of long and familiar conversations with a
friend of superior soul and mind. What subtlety and soundness of
judgment on the most diverse and most important subjects: on
education, on public administration, on everything that presented
itself to us! What freedom of mind and what gentle
cheerfulness\textemdash despite the [567:]~sufferings to which he was
already prey! He~seemed to blame himself for letting those around him
perceive~them.

He saw the approach of his end with the religious sentiments of a man
who, having been initiated further than his fellows into the secret of
the marvels of nature, was deeply permeated by the power and infinite
goodness of their Author. The services that he rendered to the
sciences by his meditations, and the useful applications that he made
of them, were in his eyes only the accomplishment of an obligatory
mission for him. It was above all by the practice of the most touching
virtues that he believed he could acquit himself before humanity, and
satisfied his conscience.

Why is this elevated soul, this happy genius, so soon taken from us?
Enduring monuments raised to the sciences, striking services rendered
to society, exemplify \textsc{Fresnel}'s too-brief existence. What did
we not still have the right to expect? Let us submit without murmur to
the irrevocable decrees of Providence; and after the initial dejection
of a grief that will follow us to the grave, let us reverently gather
and preserve the most precious of what he has left us, that which we
can try to imitate: the example of his virtues.

\begin{flushright}
\textsc{Duleau},\\
\emph{Engineer of Bridges and Roads}.
\end{flushright}

\section*{{\large Appendix:}\\Shorter obituary for Fresnel in \textit{Revue enc.}~\cite{a-1828}}\label{s-appx}

\textsc{Fresnel} (\textit{Augustin-Jean}), member of the Institute
(Academy of Sciences), born at Broglie, d\'{e}partement of Eure, on
10~May 1788, died at Ville-d'Avray, near Paris, on 14~July
1827. Admitted to the \textit{\'{E}cole polytechnique} at the age
of~16, he there distinguished himself by that patient and observant
zeal which, in the sciences, characterizes genius. Having embraced the
career of Bridges and Roads, he was successively sent by the
government, on leaving that school, to the d\'{e}partements of
Vend\'{e}e, Dr\^{o}me, and Ille-et-Vilaine, where, though entirely
faithful to the useful works entrusted to him, he found the means to
begin ingenious researches that led him to observe new facts presented
by the diffraction of light\textemdash facts inexplicable according to
the Newtonian theory of \emph{emission}, but consistent with the
theory of Huygens and Euler, which attributes the phenomena of light
\emph{to~the vibrations of a fluid spread through space}. The memoir
in which Fresnel presented the fruit of his research was sent in~1815
to the Academy of Sciences. In~1819 he~won the prize [317:]~that this
Academy had promised for the best memoir on the general phenomena of
diffraction.\footnote{\,\small Not the same memoir. Fresnel's first
memoir on diffraction~\cite{fresnel-1815}, submitted in~1815, is not
the version that appeared in print in~1816~\cite{fresnel-1816}; and
neither version is to be confused with his prizewinning
memoir~\cite{fresnel-1818mc} submitted in~1818.~\textemdash\emph{Ed.}}
Stationed in Paris by the Director-General of Bridges and Roads,
Fresnel there joined one of our most celebrated scholars, Mr.\,Arago,
continued his research, and with the aid of the principle that he had
made incontestable, managed successively to explain \emph{diffraction,
inflexion, reflection, polarization, refraction, double refraction,
etc.}\,\cite{putland-2021-}. Works of such importance opened to him
the doors of the Institute, where he was received unanimously
in~1823. He~was decorated with the Legion of Honor in~1824, and
admitted in~1825 to the Royal Society of London, which, two years
later, awarded him the prize endowed by Count Rumford, for a memoir in
which he had applied the theory of \emph{undulations} [waves] to the
phenomena of \emph{polarization}~\cite{fresnel-1821c-putland}. Since
1819 he had been attached to the Lighthouse Commission; and the
lighting of the coasts and ports owed useful improvements to him. The
exhibitions of 1823 and 1827 have shown the lenticular systems
[lenses] of Fresnel, executed by Messrs.{\tiny\!}~Soleil and
Tabouret.\, In~1823 the examining jury awarded him a gold medal and
recommended him for the Order\footnote{\,\small French:
\textit{cordon}, referring to the ribbon or lanyard of the
order.~\textemdash\emph{Ed.}} of Saint-Michel. The Society for
Encouragement [of National Industry] likewise awarded him a gold medal
in~1824.\, In~1821 he was appointed to the same examiner's post at the
\textit{\'{E}cole polytechnique} that had been occupied by Malus, who,
like him, was made famous by his beautiful observations on
light. Fresnel's naturally feeble constitution had been weakened by
the relentless continuity of his work. In~these last years his
strength declined; and after some alternations between illness and
health, he died at Ville d'Avray on 14~July 1827, aged less than forty
years, leaving a name to whose lustre a longer career would
undoubtedly have added much more.~~~\hfill\textemdash~\emph{A.}

\bigskip

\section*{Editor's acknowledgments}

A secondary goal in preparing this edition was to assess whether such
efforts are still useful in the age of AI translation, especially now
that various LLMs can translate directly from multi-page image files,
avoiding the need for manual correction of OCR errors. The first draft
of the shorter translation (appendix) was generated from a PNG image
by Google's Gemini~3. It~needed two small corrections (a~skipped word
and a quirk of French pronouns), two small clarifications (for which
it helped to know something of the subject matter), and a little
copy-editing. The first draft of the longer translation (main~text)
was generated from an image-only PDF file by OpenAI's GPT-5.1,
accessed via Perplexity~Pro, and was of comparable quality except that
the footnotes were omitted. (I~also tried Gemini on the longer
document, and Perplexity on the shorter one, with less success; but,
as the capabilities of LLMs may change from week to week, it is
probably not helpful to say more.) Having edited the first drafts with
reference to the original French, I~compared the edited text with
alternative drafts obtained from Anthropic's Claude Sonnet 4.6. There
were a few minor discrepancies, which I~again resolved with reference
to the original. \emph{Conclusion:} Machine translations (still) must
be manually checked.

Also using Claude Sonnet 4.6 (chosen by Perplexity), I~asked whether
any scholars had identified the author ``A'' of the shorter
obituary. The answer was negative, but led me indirectly and
unexpectedly to the ``Bibliographic Bulletin'' entry concerning
Duleau's original brochure (see the \hyperref[s-preface]{preface}
above). As it happened, L\'{e}onor Fresnel's obituary for
Duleau~\cite{fresnel-l-1833} was also found with the assistance of~AI
(reference \cite{pernot-23} got the year wrong). All other added
citations and references, however, are my own work.\vspace{-1em}

\begin{flushright}
\textemdash~\!\emph{Gavin R.{\small~}Putland}.
\end{flushright}\vspace{-1em}

\linespread{1}\small\raggedright

\markright{References}


\begin{thebibliography}{99}\itemsep=0.6ex

\bibitem{a-1828} A(?), ``\textsc{Fresnel}'' (obituary),\,
  \textit{Revue encyclop\'{e}dique}, vol.\,37,
  archive.org/details/revueencyclopd37pari,
  \href{https://archive.org/details/revueencyclopd37pari/page/316}{pp.\,316--7}
  (Jan.\,1828);\, translated in the above
  ``\hyperref[s-appx]{Appendix}''\!.

\bibitem{anon-1833} Anon., ``Alphonse-Jean-Claude-Bourguignon
  \textsc{Duleau}'' (obit.),\, \textit{Annales de chimie et de
    physique}, vol.\,52 (1833), archive.org/details/s3id13208210,
  \href{https://archive.org/details/s3id13208210/page/191}{pp.\,191--2}.

\bibitem{arago-1830} D.{\tiny\,}F.{\tiny\,}J.~Arago
  (tr.\,\&\,ed.~B.\,Powell), ``{\footnotesize FRESNEL}'' (elegy read
  at the Public Meeting of the Academy of Sciences, 26~Jul.\,1830),\,
  in~\!  D.{\tiny\,}F.{\tiny\,}J.~Arago (tr.~W.{\tiny\,}H.\,Smyth,
  B.\,Powell, \& R.\,Grant),\, \textit{Biographies of Distinguished
    Scientific Men} (single-volume edition), London: Longman, Brown,
  Green, Longmans, \& Roberts, 1857,
  books.google.com/books?id=m7pcAAAAcAAJ,
  \href{https://books.google.com/books?id=m7pcAAAAcAAJ&pg=PA399}
       {pp.\,399--471}.

\bibitem{bell-89} J.{\tiny\,}F.~Bell, ``Experimental solid mechanics
  in the nineteenth century'' (1989 William M.{\tiny\!}~Murray
  Lecture),\, \textit{Experimental Mechanics}, vol.\,29, no.\,2
  (Jun.\,1989), pp.\,157--65;\,
  \href{https://doi.org/10.1007/BF02321369}{doi.org/10.1007/BF02321369}.

\bibitem{boutry-48} G.A.~Boutry,\, ``Augustin Fresnel: His time, life
  and work, 1788--1827''\!,\, \textit{Science Progress}, vol.\,36,
  no.\,144 (Oct.\,1948), pp.\,587--604;\,
  \href{https://www.jstor.org/stable/43413515}{jstor.org/stable/43413515}.

\bibitem{crew-1900} H.~Crew~(ed.),~\! \textit{The Wave Theory of
  Light: Memoirs by Huygens, Young and Fresnel}~\! [vol.\,10 of
  J.{\tiny\,}S.\,Ames~(ed.), \textsc{Scientific Memoirs}],~\! American
  Book Co., 1900;\,
  \href{https://archive.org/details/wavetheoryofligh00crewrich}
       {archive.org/details/wavetheoryofligh00crewrich}.

\bibitem{duleau-1820} A.~Duleau,\, \textit{Essai th\'{e}orique et
  exp\'{e}rimental sur la r\'{e}sistance du fer forg\'{e}},\,
  Paris:~Courcier, 1820;\,
  \href{https://www.google.com/books/edition/Essai_th%C3%83_orique_et_exp%C3%83_rimental_sur/6WAUAAAAQAAJ}
    {is.gd/Duleau1820}.

\bibitem{duleau-1828} A.~Duleau, ``Notice sur A.\,Fresnel''\!,\,
  \textit{Revue encyclop\'{e}dique}, vol.\,39,
  archive.org/details/revueencyclopd39pari,
  \href{https://archive.org/details/revueencyclopd39pari/page/558}{pp.\,558--67}
  (Sep.\,1828);\, reprinted in \textit{Journal du g\'{e}nie civil, des
    sciences et des arts}, vol.\,1 (1828),
  archive.org/details/bub\_gb\_ONM-AAAAcAAJ,
  \href{https://archive.org/details/bub_gb_ONM-AAAAcAAJ/page/353}{pp.\,353--62}.

\bibitem{fresnel-1815} A.~Fresnel, ``Premier m\'{e}moire sur la
  diffraction de la lumi\`{e}re'' (submitted 15~Oct.\,1815),\,
  \textit{Oeuvres}~\cite{fresnel-oc}, vol.\,1,
  \href{https://books.google.com/books?id=1l0_AAAAcAAJ&pg=PA9}
       {pp.\,9--33}.

\bibitem{fresnel-1816} A.~Fresnel, ``M\'{e}moire sur la diffraction de
  la lumi\`{e}re''\!,~\! \textit{Annales de chimie et de~physique},
  Ser.\,2, vol.\,1, pp.\,239--81 (Mar.\,1816);\, reprinted as
  ``Deuxi\`{e}me m\'{e}moire\ldots'' (Second memoir\ldots) in
  \textit{Oeuvres}~\cite{fresnel-oc}, vol.\,1,
  \href{https://books.google.com/books?id=1l0_AAAAcAAJ&pg=PA89}
       {pp.\,89--122}. (\emph{Not}~to~be~confused
       with~\cite{fresnel-1818mc}.)

\bibitem{fresnel-1818mc} A.~Fresnel, ``M\'{e}moire sur la diffraction
  de la lumi\`{e}re'' (submitted 29~Jul.\,1818, ``crowned'' 15~March
  1819),\, reprinted (with supplements) in
  \textit{Oeuvres}~\cite{fresnel-oc}, vol.\,1,
  \href{https://books.google.com/books?id=1l0_AAAAcAAJ&pg=PA247}
       {pp.\,247--382f};\, partly translated as ``Fresnel's prize
       memoir on the diffraction of light'' in Crew~\cite{crew-1900},
       \href{https://archive.org/details/wavetheoryofligh00crewrich/page/81}
            {pp.\,81--144}.

\bibitem{fresnel-1821c-putland} A.~Fresnel
  (tr./ed.~G.{\tiny\,}R.\,Putland), ``On the calculation of the tints
  that polarization develops in crystalline plates,
  \&~postscript''\!,\, \href{https://doi.org/10.5281/zenodo.4058004}
    {doi.org/10.5281/zenodo.4058004} (2021).\, Original: ``Note sur le
    calcul des teintes que la polarisation d\'{e}veloppe dans les
    lames cristallis\'{e}es'' et seq.,\, \textit{Annales de Chimie et
      de Physique}, Ser.\,2, vol.\,17, pp.\,102--11 (May~1821),
    167--96 (Jun.\,1821), 312--15 (``Postscript''\!, Jul.\,1821);\,
    reprinted (adding section nos.) in
    \textit{Oeuvres}~\cite{fresnel-oc}, vol.\,1,
    \href{https://books.google.com/books?id=1l0_AAAAcAAJ&pg=PA609}
         {pp.\,609--48}.

\bibitem{fresnel-1821d-putland} A.~Fresnel
  (tr./ed.~G.{\tiny\,}R.\,Putland), ``Note on the remarks of Mr.\,Biot
  relating to colors of thin~plates''\!,\,
  \href{https://doi.org/10.5281/zenodo.4541332}
       {doi.org/10.5281/zenodo.4541332} (2021).\, Original: ``Note~sur
       les remarques de M.\,Biot, publi\'{e}es dans le cahier
       pr\'{e}c\'{e}dent''\!,\, \textit{Annales de Chimie et
         de~Physique}, Ser.\,2, vol.\,17, pp.\,393--403
       (Aug.\,1821);\, reprinted (adding section nos.) in
       \textit{Oeuvres}~\cite{fresnel-oc}, vol.\,1,
       \href{https://books.google.com/books?id=1l0_AAAAcAAJ&pg=PA601}
            {pp.\,601--8}.

\bibitem{fresnel-1822i-tag} A.~Fresnel (tr.~T.\,Tag), ``Memoir upon a
  new system of lighthouse illumination'' (read 29 Jul.\,1822),\,
  U.S.~Lighthouse Soc.,
  \href{https://uslhs.org/sites/default/files/attached-files/Fresnel%27s%20Memoire%20-%20Translation.pdf}
    {is.gd/Fresnel\_1822i\_Tag} (accessed 21~Feb.\,2026).\, Original:
    ``M\'{e}moire sur un nouveau syst\`{e}me d'\'{e}clairage des
    phares''\!,\, \textit{Oeuvres}~\cite{fresnel-oc}, vol.\,3,
    \href{https://archive.org/details/oeuvrescompltes01fresgoog/page/97}
         {pp.\,97--126}.

\bibitem{fresnel-1822z-putland} A.~Fresnel
  (tr./ed.~G.{\tiny\,}R.\,Putland), ``Memoir on the double refraction
  that light rays undergo in traversing the needles of quartz in the
  directions parallel to the axis'' (read 9~Dec.\,1822),\,
  \href{https://doi.org/10.5281/zenodo.4745976}
       {doi.org/10.5281/zenodo.4745976} (2021).\, Original:
       ``M\'{e}moire sur la double r\'{e}fraction que les rayons
       lumineux \'{e}prouvent en traversant les aiguilles de cristal
       de roche suivant les directions parall\`{e}les \`{a}
       l'axe''\!,\, \textit{Oeuvres}~\cite{fresnel-oc}, vol.\,1,
       \href{https://books.google.com/books?id=1l0_AAAAcAAJ&pg=PA731}
            {pp.\,731--51}.

\bibitem{fresnel-1825h} A.~Fresnel (tr.~anon.), ``On the repulsion
  exerted by heated bodies at sensible distances'' (read
  13~Jun.\,1825),\, \textit{Quarterly Journal of Science, Literature,
    and the Arts}, vol.\,20 (1826), archive.org/details/s2id13383490,
  \href{https://archive.org/details/s2id13383490/page/164}{pp.\,164--7}.\,
  Original: ``Note sur la r\'{e}pulsion que des corps
  \'{e}chauff\'{e}s exercent les uns sur les autres \`{a} des
  distances sensibles''\!,\, \textit{Annales de Chimie et
    de~Physique}, Ser.\,2, vol.\,29, pp.\,57--62,\,107--8
  (May~1825);\, reprinted in \textit{Oeuvres}~\cite{fresnel-oc},
  vol.\,2,
  \href{https://books.google.com/books?id=g6tzUG7JmoQC&pg=PA667}
       {pp.\,667--72}.

\bibitem{fresnel-oc} A.~Fresnel\, (ed.~H.{\tiny\!}~de~S\'{e}narmont,
  E.~\!Verdet, \& L.\,Fresnel),\, \textit{Oeuvres compl\`{e}tes
    d'Augustin Fresnel}~\! (3~vols.),\, Paris: Imprimerie
  Imp\'{e}riale\, (cited~as \textit{Oeuvres});
  vol.\,1,~\href{https://books.google.com/books?id=1l0_AAAAcAAJ}
  {books.google.com/books?id=1l0\_AAAAcAAJ},~1866;
  vol.\,2,~\href{https://books.google.com/books?id=g6tzUG7JmoQC}
  {books.google.com/books?id=g6tzUG7JmoQC},~1868;
  vol.\,3,~\href{https://archive.org/details/oeuvrescompltes01fresgoog}
  {archive.org/details/oeuvrescompltes01fresgoog},~1870.

\bibitem{fresnel-l-1833} L.\,Fresnel, ``{\footnotesize DULEAU}''
  (obituary),\, \textit{Association polytechnique: Compte-rendu
  trimestirel} [sic], Jan.\,1833,
  gallica.bnf.fr/ark:/12148/bpt6k6535143w,
  \href{https://gallica.bnf.fr/ark:/12148/bpt6k6535143w/f51}{pp.\,47--51}.

\bibitem{pernot-23} M.-A.~Pernot, ``{\footnotesize\bf BOURGUIGNON}
  {\bf Alphonse} Jean Claude {\bf dit} {\footnotesize\bf
    DULEAU}''\!,\, \href{https://cths.fr/an/savant.php?id=128257#}
  {cths.fr/an/savant.php?id=128257} (last modified 13~Sep.\,2023).

\bibitem{PIREN-Seine-26}
  \href{https://www.piren-seine.fr/en/le-piren-seine/who-are-we}{{\footnotesize
      PIREN} Seine}, ``{\footnotesize DULEAU}, Alphonse Jean Claude
  Bourguignon dit (1789-1832)''\!,\,
  \href{https://archiseine.metis.upmc.fr/fr/taxonomy/term/1166}
       {archiseine.metis.upmc.fr/fr/taxonomy/term/1166} (accessed
       4~Feb.\,2026).

\bibitem{putland-2021-} G.{\tiny\,}R.~Putland, ``List of English
  translations of works by Augustin Fresnel''\!,\,
  \href{https://doi.org/10.5281/zenodo.4766520}
       {doi.org/10.5281/zenodo.4766520} (2021--).\,

\end{thebibliography}
\end{document}